\documentclass[preprint2]{aastex}

\def\gapprox{\;\rlap{\lower 3.0pt                       
             \hbox{$\sim$}}\raise 2.5pt\hbox{$>$}\;}
 
\begin{document}
 
\title{Spectroscopy of Blue Stragglers and Turnoff Stars in M67 (NGC 2682)\footnote{
Based in-part on observations obtained with the Hobby-Eberly Telescope, which is a joint project of 
the University of Texas at Austin, the Pennsylvania State University, 
Stanford University, Ludwig-Maximillians-Universit\"at M\"unchen, and
Georg-August-Universit\"at G\"ottingen.}}

\author{Matthew D. Shetrone}
\affil{University of Texas/McDonald Observatory,
P.O. Box 1337, Fort Davis, Texas 79734}
\email{shetrone@astro.as.utexas.edu}

\author{Eric L. Sandquist}
\affil{San Diego State University,Department of Astronomy,San Diego, CA 92182}
\email{erics@mintaka.sdsu.edu}

\begin{abstract}

We have analyzed high-resolution spectra of relatively cool blue
stragglers and main sequence turnoff stars in the old open cluster M67
(NGC 2682).  We attempt to identify blue stragglers whose spectra are
least contaminated by binary effects (contamination by a binary
companion or absorption by circumstellar material). These ``best''
stragglers have metallicities ([Fe/H] = --0.05) and abundance ratios
of the blue stragglers are not significantly different from those of
the turnoff stars. Based on arguments from hydrodynamical models of
stellar collisions, we assert that the current upper limits for the
lithium abundances of all blue stragglers observed in M67 (by us and
others) are consistent with no mixing during the formation process,
assuming pre-main sequence and main sequence depletion patterns
observed for M67 main sequence stars.  We discuss composition
signatures that could more definitively distinguish between blue
straggler formation mechanisms in open cluster stars.

We confirm the spectroscopic detection of a binary companion to the
straggler S 1082.  From our spectra, we measure a projected rotational
speed of $90\pm20$~km s$^{-1}$ for the secondary, and find that its
radial velocity varies with a peak-to-peak amplitude of $\gapprox$ 25
km s$^{-1}$. Because the radial velocities do not vary with a period
corresponding to the partial eclipses in the system, we believe this
system is currently undergoing mass transfer. In addition we present
evidence that S 984 is a true blue straggler (and not an unresolved
pair). If this can be proven, our detection of lithium may indicate a
collisional origin.

\end{abstract}

\keywords{stars: abundances --- stars: blue stragglers --- open
clusters: individual (NGC2682, NGC7789)}

\section{Introduction}
 
Blue straggler (BS) stars were first identified as an unusual subclass of
stars in the cluster M3 \citep{sandage53}. Since that time, stragglers
have been identified in most, if not all, globular clusters that have
accurate broad-band photometry. Blue straggler candidates have also
been identified in a large number of open clusters, e.g. \citet{ahumada95}.

Blue stragglers are commonly defined as those stars that
are brighter and bluer (hotter) than the main sequence (MS)
turnoff of the majority of the cluster stars, but along an apparent
extension of the main sequence. At present, the leading explanation
involves mass transfer in and/or merger of a binary star system, or the
collision of stars (whether or not a binary is involved). These 
mechanisms can create a main sequence star with a mass greater
than would be expected given the age of the cluster.

The above definition of blue straggler is overly restrictive if
this is the correct explanation. Undoubtedly the same processes
act on lower mass stars, although the remnant stars would not be
easily identifiable if they were not brighter than the turnoff
luminosity. The color restriction is also unrealistic since real
scatter in the colors of blue stragglers in globular clusters has been
observed, e.g. \citet{fusipecci92}. \citet{sills99} find that this
color scatter results from a lack of large-scale mixing during the
creation process, which means that the amount of helium in the core
can vary from straggler to straggler.

Hydrodynamical studies \citep{lombardi95,sandquist97} earlier
predicted that large-scale mixing does not occur during stellar
collisions. A collisionally-produced straggler is not expected to burn
significant lithium initially even if high enough temperatures are
reached because the timescale for the straggler to thermally adjust to
a new equilibrium configuration is much shorter than the lithium
destruction timescale \citet{sills97}. Binary mass transfer is likely
to lead to a more easily observable abundance pattern since the last
of the transferred gas should have come from deep within the parent
star \citep[e.g,][]{sarna96}.

In this paper we have attempted to examine this question by taking
spectra of blue straggler stars in the open cluster M67 (NGC 2682).
M67 makes an ideal target because it is relatively close
\citep[$V-M_{V}\approx 9.7$;][]{dinescu95}, and also fairly old
\citep[$\sim 4.0$~Gyr;][]{dinescu95}. In addition, the density of
stars in open clusters is lower than in nearly all globular clusters,
making binary evolution and collisions involving binaries the most
likely formation mechanisms \citep{leonard92}. Globular cluster blue
stragglers are not yet practically observable because of their
distances. Although much attention has been spent on M67 blue
stragglers, few observations are available for the coolest ones.

Few blue stragglers have abundance analyses.  \citet{mathys91}
observed a sample of 11 blue stragglers in M67, although only 2 were
examined in detail. \citet{andrievsky98} observed 4 blue stragglers in
Praesepe (NGC 2632) from the catalog of \citet{ahumada95}, although
only one (HD 73666) is undeniably a straggler according to its
photometry. In observing the blue stragglers of M67 we are most
interested in species such as Li, C, N, and O that are processed by
nuclear reactions at different temperatures (and therefore, depths) in
a star. The reddest blue stragglers in the cluster have effective
temperatures that can allow us to get abundances for all of these
elements, and are close enough to the temperatures of turnoff stars
that we can straightforwardly make comparisons with stars having
(presumably) unmixed envelopes.

In \S 2 we describe the spectroscopic observations, and in \S 3 we present
the abundance analysis for our sample. In \S 4 we discuss the constraints
we can place on blue straggler formation mechanisms in this cluster.
 
\section{Observations}

Table~\ref{tbl-1} displays the program stars, basic photometry, and
the signal-to-noise obtained. The identification numbers we use are
from the proper motion study of \citet{sanders77}. We chose our blue
straggler candidates from \citet{ahumada95} and our MS members from
the proper motion study of \citet{zhao93}.  All of the chosen M 67
stars are 99\% probability proper motion members.  The photometry for
our program stars was taken from \citet{fan96}, \citet{montgomery93},
\citet{chevalier91}, \citet{gilliland91}, and \citet{sanders89}.  The
$B-V$~ colors of \citeauthor{sanders89} and \citeauthor{gilliland91}
were adjusted by 0.006 and $-0.002$, respectively.  The $V$ magnitudes
of \citeauthor{sanders89}, \citeauthor{gilliland91} and
\citeauthor{chevalier91} were adjusted by 0.017, $-0.009$, 0.012,
respectively.  Several measurements made by \citeauthor{sanders89} were
found to be in disagreement with the other sources, including
ancillary measurements by \citet{anupama94}, \citet{murray65}, and
\citet{murray68}.  For this reason, the \citeauthor{sanders89} measurements
for S 1183, S 821, S 975, and S 984 were not used to derive the
average photometry.  The final adopted values are given in
Table~\ref{tbl-1}, and the CMD of the cluster and the observed blue
stragglers are shown in Figure 1.  We observed one blue straggler
candidate \citep{ahumada95} in NGC 7789.  Due to poor weather
additional NGC 7789 candidates were not observed.

The majority of the observations were taken with the McDonald
Observatory 2.7~m "2d-coude" \citep{tull95}.  The resultant spectra
have $R=30,000$ and cover from 3800 \AA ~to 10100 \AA.  
Each star was observed on several different nights
in order to look for radial velocity variations.  Radial velocities
were obtained from individual spectra, but the S/N quoted in
Table~\ref{tbl-1} are those of the combined spectra.  The combined
spectra were also used for the abundance analysis.  Additional
observations were obtained with the Hobby-Eberly Telescope 
\citep[HET]{ramsey94}
during commissioning of the UFOE \citep{harlow96} 
spectrograph in early 1999 and 
in the first science period of 2000 with the UFOE
on the HET.  The HET spectra were used to obtain radial velocities and
have lower resolution ($R=11,000$) than the 2.7~m spectra, and so are
not included in the combined spectra.  The HET spectra cover from
4600 \AA ~to 9100 \AA.  In March 2000, the McDonald
Observatory 2.1~m Cassegrain echelle was used to obtain a few
additional observations of two of the program stars.  Despite having
resolution higher than the 2.7~m "2d-coude" spectra, these were not
included in the abundance analysis because the slit length did not
allow proper sky subtraction, which would not
interfere with the radial velocity study since M67 is well separated
from the solar rest velocity.  The 2.1~m spectra cover from 
5400 \AA ~to 6600 \AA.   Table~\ref{tbl-2} lists the
heliocentric Julian dates of our observations.  In addition to the
program stars one additional M67 star (MMJ5554) was observed.  This
star is only $3^{\prime \prime}$ away from S 975 and its color makes
it either a turnoff star or a possible blue straggler candidate.  A
set of bright, rapidly-rotating hot stars were observed at various air
masses with both telescopes for sky line identification.  Figure 2 
exhibits a sample of the spectra obtained on the McDonald 2.1m.

\section{Analysis}\label{analysis}

\subsection{Radial and Rotational Velocities}

The spectra were analyzed using the IRAF {\it echelle} package.
Initial data reduction techniques included bias removal, flat field
correction, scattered light removal, sky subtraction and wavelength
calibration.  To remove instrumental wavelength zero-point errors the
spectra were shifted in velocity so that the telluric features fell at
their rest velocities.  The largest zero-point correction was 1.1 km
s$^{-1}$.  Each program star spectrum was divided by the hot star
spectrum with the most similar telluric line depths.  The spectra were
then cross-correlated against a synthetic spectrum to obtain the
radial velocities, which were then corrected to the heliocentric rest
frame.  A single synthetic spectrum was used for all the stars
($T_{eff} =$ 6400 K, log g = 4.3, $v_t$ = 2.0, [Fe/H] = --0.05, and no
instrumental or rotational broadening).  The code used to create the
synthetic spectrum is the latest version of MOOG \citep{sneden73}.  To
determine the radial velocities of the broad component of S 1082 we
took an additional step to remove the narrow component.  We fit the
spectrum with a very high order spline (high enough to remove any
broad component) and then divided this normalized spectrum into the
original spectrum leaving only the blaze function and any broad line
components.  These cleaned broad-line spectra were then subjected to
the analysis described above.  Table~\ref{tbl-2} lists the
heliocentric radial velocities and errors for each spectrum.  The mean
radial velocity of M67 is 33.6 $\pm$ 0.72 km s$^{-1}$
\citep{girard89}.  We detect 3 binary periods not previously known for
stars in our sample.  S 821 and S 984 are found to have radial
velocity variations with periods of 26.259 $\pm$ 0.002d and 1.465
$\pm$ 0.001, respectively.  Figure 3 shows the phased radial velocity
data from \citet{mathieu86} and this work for these periods.  S 821
probably has an eccentricity of $0.4\pm0.1$, while the data from S 984
is consistent with a circular orbit.  S 1082 is found to be a double
lined spectroscopic binary and will be discussed in \S 3.

The 2.7~m "2d-coude" spectra were combined after shifting to the heliocentric 
rest velocity.  New synthetic spectra were created with $T_{eff} =$ 6400 K,
log g = 4.3, $v_t$ = 2.0, [Fe/H] = --0.05, a Gaussian $R=30,000$ profile,
and various rotational velocities.  The unbroadened spectrum was
cross-correlated against the broadened spectra and the full width half
maxima (FWHM) of the correlation peaks were measured. These FWHM were
then compared to the FWHM of the correlation peaks created by the
cross-correlation of the combined program star spectra and the
unbroadened synthetic spectrum.  From this comparison we determine the
rotational velocity of the combined spectrum.  We adopt a conservative
lower limit (9 km s$^{-1}$) to our ability to measure rotational
velocities.  Table~\ref{tbl-1} lists the rotational velocities found
with this method.

We detect rotation in our spectra for only 3 of the blue straggler
stars in our sample: S 975, S 997, and S 1082. All have relatively low
rotation speeds. In addition, we marginally detect rotation in one of
the main sequence stars (S 1271). The signal-to-noise ratios for the
main sequence stars was comparable to those for the blue stragglers.

We observed one blue straggler candidate in NGC 7789, M 1251, but
found it to be a cluster non-member based on its radial
velocity.

\subsection{Abundance Analysis}

The spectral lines used in our analysis were taken from
\citet{shetrone96}, \citet{mathys91}, and \citet{edvardsson93}.  The
oscillator strengths were taken from these sources as well as the
National Institute of Standards and Technology Atomic Spectra
Database.  Each line's equivalent width was measured with the IRAF
task {\it splot}.  These lines, oscillator strengths, and equivalent
widths are listed in Table~\ref{tbl-3}.  We employed the latest
version of the LTE spectral analysis code MOOG \citep{sneden73} and
the \citet{kurucz93} grid of ATLAS models in an iterative abundance
analysis.

The initial estimate of the effective temperature for each star was
determined from the dereddened $B-V$ color (we assumed E$(B-V) =
0.05$).  Effective temperatures were then determined from the
relationship $T_{eff} = 1808(B-V)^2 - 6103(B-V) + 8899$
\citep{soderblom93}.  Each effective temperature was then fine tuned
by forcing the slope of abundances from Fe I lines versus excitation
potential to be zero.  This tuning of the temperature was restricted
to the range of colors consistent with the errors in the photometry.
The two largest corrections were found for S 1082 (152 K; likely
binary blending of colors, see \S 3) and S 815 (42 K).  The surface
gravity was set by enforcing abundance equilibrium between the Fe I
and Fe II lines.  Based upon the errors in the Fe I and Fe II abundances
we estimate the average uncertainty in the surface gravity to be
0.15 dex (1 sigma).  The microturbulent velocity was determined by
forcing the slope of Fe I line abundances versus equivalent width to
be zero.  These adjustments to the spectral model for each star were
done iteratively until all the parameters ($T_{eff}$, log g, $v_t$)
fulfilled the above requirements within the abundance uncertainty of
each spectrum.  The derived atmospheric parameters and abundances for
each star are listed in Table~\ref{tbl-4}.

Synthetic spectra were also employed to verify the abundances of the C
I lines.  Using the derived atmospheric parameters for each star a
synthetic spectrum was generated and plotted over the observed
spectrum and visually inspected.  This method did not yield any
significant abundance differences.

The three O I lines found at 7773\AA ~are known to need NLTE
corrections.  These corrections can become very significant at high
temperatures (T $>$ 6800 K).  Over the narrow range of temperatures we
considered the NLTE corrections should be relatively small.  However,
we have made some attempt to correct for NLTE effects using the lookup
table employed by \citet{gratton99}.  Even if the magnitude of our
corrections are incorrect, the relative oxygen abundances of the stars
can be viewed with some confidence.  In Figure 4 we display the
combined equivalent widths of the O I lines plotted against their
colors.  The stars from \citet{mathys91} are shown as squares.  The
large crosses are taken from the \citet{varenne99} study of oxygen in
the Hyades.  \citet{varenne99} found a constant abundance along the
main sequence.  Note the rapid rise in O I EW near $B-V = 0.43$ or
$T_{eff} =$ 6840 K, caused by the onset of NLTE effects.  The open and
filled circles represent abundance analyses with constant oxygen
abundance ([O/Fe] = 0.0) for several temperatures with the NLTE
corrections of \citet{gratton99} and \citet{faraggiana88},
respectively.  A summary of our oxygen abundances is given in Table 4.

\section{Discussion}
 
Because the stragglers in M67 have been observed extensively, we have
a better idea of their present status than the stragglers in any other
cluster.  It is therefore worth briefly summarizing what is known
about the stars in our sample so that we can test whether the
compositions reflect differences in straggler formation mechanisms.

\subsection{Discussion of Individual Stars}

\subsubsection{Blue Stragglers}

{\bf S 975 (F 90):} This star is known to be a binary ($P = 1221$~ d)
with small eccentricity ($e=0.088\pm0.060$) from the work of
\citet{latmil96}. \citep{leonard96} indicates that the system is
likely to have resulted from case C (asymptotic giant branch) mass
transfer. \citet{landsman98} assert that there is strong evidence of a
hot companion in their ultraviolet photometry, and conclude that the
system is likely to be the result of Algol-type binary mass transfer,
although they adopt a photometric temperature
\citep{mathys91} that is more than 300 K cooler than our derived
value. \citeauthor{mathys91} states that the photometry could be
contaminated by a nearby star, so that his temperature estimate is
questionable. The contaminating star must be cooler than S 975 in
order to produce the photometric temperature. MMJ 5554 ($V = 12.839,
B-V = 0.532$) is probably the cause since it is only $3^{\prime
\prime}$ away from S 975, and its colors place it very near the bluest
extension of the turnoff. Because our spectral temperature did not
need to be significantly corrected from our $BV$ photometric
temperature, it is unlikely that this star contaminated our
spectrum. Judging from Figure 2 of \citet{landsman98}, it probably
did not contribute significantly to the UIT UV flux.

According to \citeauthor{landsman98}, ``uncertainty of 200 K in the
value of $T_{eff}$ corresponds to about a 0.5 mag uncertainty in the
predicted UIT magnitude'', and ``the UIT magnitude of F90 is probably
uncertain by close to a factor of 2'' because it was contaminated by
the light of the bright blue straggler S 977. Based on this and our
higher $T_{eff}$, the evidence for a {\it hot} companion to S 975 is
considerably weaker, although there is still an unaccounted-for UV
excess. The agreement between our photometric and spectroscopic
temperatures indicates that the companion does not contribute
significantly to the visual flux (consistent with being a white
dwarf). However, we have been unable to detect the companion's
spectral signature, so we are unable to judge whether the spectrum is
contaminated.  The star does have an unusually large O I EW though
(see \S \ref{CNO}), which may betray the presence of circumstellar
material.

{\bf S 984 (F 134):} Because S 984 lies within 0.75 mag of the
fiducial line of the cluster in the color-magnitude diagram, it is
possible that it is a non-interacting binary composed of two main
sequence stars. The velocity dispersion in the radial velocity data of
\citet{mathieu86} was marginally higher than average.  After combining
the velocity data from \citet{mathieu86} with our derived velocities, we
conducted a periodicity analysis and found a likely period
of 1.465 $\pm$ 0.001 days. The peak-to-peak variation is between about
3 and 5 km s$^{-1}$, depending on whether we accept one outlying observation
from Mathieu et al.

The period, radial velocity variation, and photometry do constrain the
identity of S 984 though. If we hypothesize that S 984 is a binary
composed of stars that have not interacted, it is possible to match
the photometry with two main sequence stars having $V \approx 12.7$
and 13.4. This is the maximum brightness contrast that is possible if
the primary is to be on the fiducial line of the cluster. However,
this constrains the mass ratio to be $0.9 < q \leq 1.0$. The orbital
inclination is then constrained to be within 1.5 degrees of
face-on. The lithium abundance, which is consistent with that of a
single turnoff star, is low for a tidally-locked binary (as is implied
if the radial velocity variations reflect orbital motions).  So, there
is a fairly strong (although circumstantial) case that S 984 is a true
blue straggler, and not a photometric blend of two stars. There may
still be a companion, but it is unlikely to contribute to the
flux. If this is
the case, S 984 is probably the result of a collisional merger, since
binary mass transfer should result in a lack of surface lithium. For
these reasons, we regard it as likely that the spectrum of S 984 is at
most minimally contaminated by another star.

{\bf S 997 (F 124):} This star is known to be in an eccentric
($e=0.342\pm0.082$) binary of period 4913 d \citep{latmil96}. The
eccentricity of the system probably rules out a history as a mass
transfer binary.  The fact that the primary is a blue straggler
casts doubt on a binary merger since it would either have required the 
system to have been in a rare triple system, or to have tidally
captured a star after merger \citep{leonard96}. Either way, we do not
have evidence to judge whether its spectrum is contaminated.

{\bf S 1082 (F 131; ES Cancrii):} \cite{pritchet91} and
\citet{mathys91} reported that the spectrum of S 1082 has a possible
composite nature. X-ray emission has been detected from this star by
\citet{belloni93}. \citet{landsman98} find evidence of a hot
subluminous companion in their ultraviolet photometry.  However,
\citet{landsman98} base their predicted UIT magnitudes on a
temperature 120 K cooler than found in our analysis.  Since a 200 K
change in temperature corresponds to 0.5 m$_{152}$ magnitudes, their
evidence for a {\it hot} subluminous companion is weakened, although
not eliminated. The discovery of a variable secondary
H$\alpha$ spectral feature by \citet{mathys91} and \citet{vdb99}
suggests the presence of a hot rapidly rotating secondary or an
outflowing wind.  Further, the spectroscopy of \citet{vdb99} appears
to rule out magnetic activity as the source of the X-ray
emission. This evidence supports the hypothesis that this is an
Algol-type binary {\it currently} undergoing mass transfer.

In addition to being able to detect the second spectral component in
H$\alpha$, the Na D lines, and the O I triplet as \citet{mathys91}
did, we also detected the secondary in H$\beta$, H$\gamma$,
Paschen(8862), the Ca II IR triplet, lines of moderately strong
neutral element species (e.g. Fe I, Ca I) and a few strong Fe II
lines.  The abundances derived for 1082 are 0.2 dex lower than the
cluster mean.  This can be explained if the second component produces
40--50$\%$ of the continuum flux but only a tiny ($\sim$10$\%$)
contribution to the line flux. The rotational velocity implied for the
second spectral feature is 90$\pm$20 km s$^{-1}$, which explains the
small contribution to the line flux.

For the first time, we are able to see radial velocity variations in
the secondary (broad-lined) component amounting to at least 25 km
s$^{-1}$.  Our phase coverage is not good enough to state more than a
lower limit. 
Figure 2 exhibits the series of spectra we obtained of 
S 1082 on the McDonald 2.1m.   The broad component can clearly be
seen when compared with the spectra of S 984.   In addition, the 
broad component of S 1082 moves redward over the 5 hours the star
was observed.
For the primary, we find a peak-to-peak amplitude of
about 5.5 km s$^{-1}$. From the data of \citep{mathieu86}, there is a
peak-to-peak amplitude of about 7 km s$^{-1}$. 

\citet{goranskij92} found from photometry that the system has two
partial eclipses per orbit of different depth, with a period of
1.0677978 d. Using this period (or half of it), we see no sign of a
repeating pattern in the radial velocity data. Allowing the period to
vary, we find a highest probability period of 1.87 $\pm$ 0.05 d, with
less likely values of 0.7 d and 2.3 d (see Figure 3). However, none of these
periods unveils a satisfactory pattern in the radial velocity data.
We have two
sequences of spectra taken on succeeding nights (centered around phase
0.5 of the photometric period) that indicate sudden ($\approx 2$ h)
jumps in radial velocity for both components. Thus, the radial velocity data
imply that we are seeing the signature of material outside of the
photospheres of the two stars.

The spectral temperature derived for the narrow component is more than
100 K hotter than the photometry predicts.  Using the color as the
composite temperature for the system, our spectroscopic temperature
suggests that the narrow line component may be 200 K hotter than the
broad lined component.  Further evidence for a temperature difference
between the two components comes from the spectrum: the broad
component's high excitation lines of O I have smaller equivalent
widths than those of the narrow component, while the low excitation Na
D lines have higher equivalent widths. This implies a cooler
temperature, although strong Fe II lines may indicate that a lower
surface gravity also comes into play. The relatively large brightness
(M$_V \sim 2.5$) and hot temperature ($T \sim 6800$ K) of the
secondary indicates that it has probably been severely disturbed by
its interaction with the primary.

{\bf S 2204 (F 130):} We observed this star initially believing it to
be a main sequence star. Evidence indicates that it is very probably a
faint blue straggler, and it is identified as such in the catalog of
\citet{ahumada95}. The proper motion studies of \citet{girard89} and
\citet{sanders77} cite membership probabilities of 98\% and 83\%
respectively, and radial velocity measurements by \citet{mathieu86}
and us put it right on the cluster average. The photometry of
\citet{montgomery93} puts it ($B-V=0.448$) over 0.07 mag to the blue
of the {\it bluest} stars at the turnoff, and over 0.1 mag from the
fiducial line of the cluster. The quoted mean error in the photometry
was 0.016. Significant contamination of the spectrum by a main
sequence companion is not likely since this would either make the star
brighter than the turnoff, or closer in color.

A hot, faint companion could also potentially create the observed
colors. Although S 2204 was in the field of the ultraviolet
observations of \citet{landsman98}, it was not detected. Their
detection of several white dwarf candidates thereby places an upper
limit on the temperature of a possible white dwarf companion.
\citet{landsman98} estimate their detection threshold for a white
dwarf with a pure hydrogen atmosphere would be about 21,300 K.  A
white dwarf below the detection threshold would change the broadband
color by a negligible amount.

For stars close to the turnoff, the possibility of delayed star
formation (a very unlikely possibility for brighter blue stragglers)
should be considered.  For S 2204, a delay of more than 1 Gyr (about
25\% of the cluster's age) would be necessary according to the
isochrone fits of \citet{dinescu95} if S 2204 is an undisturbed main
sequence star. On this basis we discard the delayed star formation
picture for S 2204.

With the elimination of these hypotheses, the most likely
explanation of S 2204 is as a blue straggler of the merged-star or
mass-transfer variety. Although there is currently not enough radial
velocity information to more strongly test for the presence of a
companion to S 2204, we argue that the primary would be identified as
a blue straggler even without one, and that such a companion would 
not significantly modify the spectrum of the straggler.

\subsubsection{Turnoff Stars}

{\bf S 821:} We have detected binary motions for this object with a
period $P = 26.259 \pm 0.002$ d and eccentricity $e \approx 0.4 \pm
0.1$ ({\bf Figure~3}).  Broad-band photometry places it very close to
the fiducial line of the cluster.  This probably indicates that
the spectrum is uncontaminated by the companion, although there is a
chance that it is composed of two fainter, nearly equal mass stars.

\subsection{Composition Analysis}

The blue stragglers we have observed can be divided into two {\it
probable} categories: mass transfer binaries (S 975, S 1082), and
collision products (S 984, S 997). At present we do not have enough
information to classify S 2204. We also have spectra for 5 stars near
the main sequence turnoff (S 815, S 821, S 1183, and S 1271). Our goal
is to compare the abundances to see if we can find chemical signatures
of the blue straggler formation process.

\subsubsection{Lithium}

If significant mixing of a star's envelope has occurred, then the
lithium abundance may show the results first since it is consumed at
relatively low temperatures ($T \gapprox 2.5 \times 10^{6} K$).
\citet{jones99} present the most recent compilation of lithium
abundances for M67 stars.

Our results are given in Table~\ref{tbl-4}. Two of our stars (S 997
and S 2204) were previously observed by \citet{garcia88}, who found
upper limits. We derive upper limits that are 0.5 and 0.8 dex lower,
respectively. With the exception of S 984, we are only able to derive
upper limits for the stars in our sample. The values for the
main sequence stars are consistent with the lower envelope of lithium
abundances \citep[see][for references to other papers on M67 lithium
abundances]{jones99}.

\citet{hobbs91} previously observed the relatively cool blue stragglers
S 1072 and S 1082, while \citet{pritchet91} observed stragglers S 752,
S 997, S 1082, S 1263, S 1267, S 1280, and S 1284. Both sets of
authors found only upper limits for their samples. So, the lithium abundances
do not as yet provide a dependable way of distinguishing among different
formation mechanisms.

\subsubsection{CNO elements}\label{CNO}

The CNO elements can potentially constrain the amount of mixing that
occurs in a star, since the CNO cycle starts to shuffle abundances at
around $10^{7}$ K. Surface material is unlikely to reach this
temperature in main sequence stars in the mass range present in M67,
even if violent interactions occurred. However, if mass transfer were
to occur on the giant branch, processed material might be dropped onto
the surface of the straggler.

Our results for carbon and nitrogen are given in Table~\ref{tbl-4},
while our oxygen results are found in Table~\ref{tbl-5}. Our nitrogen
abundances are very uncertain because of the very small equivalent
widths found at this temperature, so we will not discuss them further
except to note that the nitrogen abundances for the blue stragglers
appear to be lower (0.2 -- 0.3 dex) than the abundances found in the
turnoff stars.  The carbon abundances for all of the stars in our
sample fall very close to the solar ratio.  Comparing the oxygen
abundances of the blue stragglers as a group to those of the main
sequence stars, the blue stragglers (dropping S 975) also have an
average abundance near that of the turnoff stars. However, several stragglers 
have anomalously high or low values.

The straggler S 975 is problematic to analyze. On one hand, the
evidence points to mass transfer of the type that could produce an
observable chemical signature. On the other hand, the composite nature
of the system appears to be affecting our equivalent width
measurements.  The measured O abundance is rather high compared to all
of the other stars in our sample. The expectation is that the mass
transfer should lead to a depletion of oxygen if there was CNO
processing. The anomalous abundance is possibly due to excess line
flux contributed by a companion star or to inaccurate NLTE
abundance corrections since S 975 lies very near the point where the
NLTE corrections become large.

For the two stars (S 975 and S 1082) that overlap with the blue
straggler sample of \citet{mathys91}, the agreement of the O I triplet
equivalent widths is very good. We have attempted to test the
hypothesis that the blue stragglers and main sequence stars observed
all have the same oxygen abundance. As seen in Figure 4, the majority
of the stars have EWs that are consistent. The exceptions are S 975
(which appears to have a companion that affects the line flux),
S 1434 (B--V = 0.12, EW(O I) = 1071 \citet{mathys91})
\citep[which also has a high EW, possibly related to a
post-main sequence companion inferred from infrared
photometry;][]{peterson84}, and S 968 
(B--V = 0.12, EW(O I) = 687 \citet{mathys91}).
S 968 is known to be an Am star, which may
account for the difference in that case. However, S 752 
(B--V = 0.29, EW(O I) = 754 \citet{mathys91})
is also an Am
star, and it does not appear to have an anomalous equivalent width.
Finally, we find that S 1082 has a low O EW because of the continuum flux
contribution of an nearly equal brightness secondary.

To aid in the interpretation of this figure we compare stars from
other studies.  The large crosses are taken from the \citet{varenne99}
study of oxygen along the main sequence of the Hyades.
\citeauthor{varenne99} found a nearly constant abundance along the
main sequence.  Note the rapid rise in O I EW at $B-V = 0.43$
($(B-V)_{0} = 0.38$) or $T_{eff} =$ 6840 K due to NLTE effects.  The
open circles and filled circles represent synthetic abundance analysis
employing constant oxygen abundance ([O/Fe] = 0.0) for several
temperature with the NLTE corrections of \citet{gratton99} and
\citet{faraggiana88}, respectively.
 
To the accuracy of our measurements, we see no evidence for
differences between the abundances of the blue stragglers and the
turnoff stars.

In an analysis of two hotter blue stragglers in M67, \citet{mathys91}
found total CNO abundances that were lower than those of giants in the
cluster.  Carbon and oxygen in particular were significantly depleted
relative to iron. None of the blue stragglers in our sample show this
pattern of depletion. Nor do our blue stragglers show differences in C
and O abundances compared to our turnoff stars. The abundance results
are thus consistent with a collisional formation. However, the low
eccentricities of some of the binaries containing stragglers are
difficult to explain in that case.

\subsubsection{Other Elements}

Nuclear processing near the hydrogen burning shell of a giant star
could produce anomalous abundances in other, more easily observable
elements.  Surface anomalies in globular cluster giant stars have been
observed in a number of globular clusters (e.g. M3, M13, M15, M71,
M92, 47 Tuc).

\citet{gilroy91} showed that the carbon isotope ratio
C$^{12}$/C$^{13}$ in M67 giants is lower than can be predicted by
standard evolutionary theory. However, the required extra mixing does
not extend into regions where the oxygen abundance is modified, as is
found in metal-poor globular clusters. In fact, the amount of mixing
seen in disk giants \citep{lambert81} is small in comparison to
globular cluster giants \citep{shetrone96}.  Na, an element which is
burned at a relatively low temperature \citep{langer97,cavallo98}, 
has not been studied well in open cluster giants.

Here in particular our ability to discern which stragglers are most
likely to be free from contamination is important. The best candidates
(S 984 and S 2204) show fairly minor differences in abundance with the
turnoff stars.  We must also be aware that there may also be selection
effects in that stragglers that form by some mechanisms may have
unavoidably contaminated spectra.  The straggler S 1082 shows
abundances that set it off from the other stars in our sample --- in
particular Ni and Ba have lower than expected abundances. The measured
iron abundance is {\it very} different from all of the other cluster
stars. The temptation is to think that this may relate to the mass
transfer that is inferred.  But as mentioned above this star is part
of a binary having a composite spectrum that is difficult to
disentangle.

\subsection{Interpretation}

\subsubsection{Lithium}

To date there has not been a detection of lithium in 
a blue straggler in an open cluster. There are several possible reasons 
for this, some of which have interesting consequences.

If two stars were to mix completely during a collision, depletions of
a factor of 50 might be expected \citep{pritchet91} because lithium
would get diluted throughout the blue straggler. However,
hydrodynamical studies of stellar collisions
\citep{lombardi95,sandquist97} indicate that the collision itself is
unlikely to mix the stars substantially --- the high entropy gas near
the surfaces of the input stars remains at high entropy near the
surface of the remnant \citep{lombardi96}.  As a result, a
collisionally produced straggler is not expected to burn significant
lithium initially, even if high enough temperatures are reached
\citep{sills97}.

For blue stragglers at or above the turnoff of the cluster that were
formed via stellar collisions, the lithium abundance observed at the
surface is therefore likely to be the same as the abundance at the
surface of the input star with the higher surface entropy. For main
sequence stars, this would be the more massive star. If the difference
in mass of the input stars is relatively large, gas from the lower
mass star is likely to be completely sequestered in the interior of
the straggler where it would be unobservable.

For open clusters, it is well known that lithium abundance drops with
decreasing effective temperature along the main sequence
\citep[e.g.][]{jones99}. For an open cluster as old as M67, two-body
collisions that produce an identifiable blue straggler could involve a
star of as low a mass as 0.6 M$_\odot$ as the more massive star. According
to standard stellar models, main sequence stars with masses less than
about 0.8 M$_\odot$ are likely to have burned enough of their lithium that
it would be undetectable. Standard models are currently unable
reproduce the lithium abundance patterns in the Hyades
though. Extrapolating from Hyades measurements, stars with masses less
than about 1.0 M$_\odot$ are likely to be undetectable.

While collisions of 0.3 M$_\odot$ and a 1.0 M$_\odot$ stars could
create blue stragglers with observable lithium, two factors work
against this.  First, such collisions are relatively rare thanks to
the bias in the cluster luminosity function toward stars relatively
near the turnoff \citep{montgomery93}.  Second, significant main
sequence Li depletion {\it is} possible after the blue straggler is
formed. In the Hyades, a dip in lithium abundances is observed for
stars with $T_{eff} \approx 6600$~K (the total range is about 300
K). It is believed that this feature may be due to meridional
circulation or shear turbulence \citep[for a recent review,
see][]{talon98}. For M67 no current main sequence members fall in this
region, but some blue stragglers do.  S 997 and S 2204 fall in the Li
gap and may have depleted their Li even if they arrived on the blue
straggler sequence with a significant supply.  The age of the Hyades
\citep[$\sim 625$ Myr;][]{perryman98} provides an upper limit for the
lithium destruction timescale for these stars if this mechanism does
act. The lifetime of a blue straggler falling in that range of
effective temperatures would be about 3 Gyr, in which case there is
less than 25\% probability of finding lithium. Blue stragglers which
form with more helium in their cores (because the lower mass input
star was relatively evolved) would have higher probabilities of
showing lithium, but shorter lifetimes as blue stragglers.

More massive blue stragglers ($T_{eff} \gapprox 7000$~K) would be more
likely to retain surface lithium because both the straggler and the
input stars lack extensive surface convection zones, and because the
stragglers are hotter than the lithium gap. A possible problem is the
mass loss associated with a collision. In hydrodynamical models of
globular cluster star collisions by \citet{lombardi96} and
\citet{sandquist97}, this mass loss was less than approximately 6\% of
the total input mass, with the largest mass loss occuring for
relatively rare head-on collisions. However, mass loss would tend to
remove the most lithium-rich material (for example, see Table 4 of
\citet{lombardi96}). For stars with masses between about 1.0 and 1.25
M$_\odot$ (approximately the turnoff mass), the lithium-rich regions
of the star add up to approximately 4-5\% of the mass.

Blue stragglers formed via binary mass transfer are unlikely to show
surface lithium \citep{sills97}. In principle, the brighter, more
massive blue stragglers are the ones to observe to look for lithium
evidence that different mechanisms are responsible for the formation
of M67 blue stragglers.  Unfortunately, the Li line is very temperature
sensitive and becomes extremely weak in hot massive stars.

\subsubsection{CNO elements}

Although they are less likely to show the effects of relatively low
levels of mixing, CNO elements may prove to be more useful in
distinguishing between stragglers that have formed by collisions and by binary
mass transfer. Because the CNO elements undergo nuclear processing at
higher temperatures than lithium, it is very unlikely that the surface
abundances will be modified as the result of a collisional merger.
Massive input stars will not have modified their surface CNO abundances
because their convection zones are not extensive enough, while low mass input
stars will in general become completely engulfed in a higher mass star due
to their lower entropies.

On the other hand, if Algol-type binary mass transfer has occurred (on
the red giant branch or asymptotic giant branch), the mass visible at
the surface of the blue straggler is taken from deep
inside the primary where it could have undergone nuclear
processing. Spectroscopy of giant branch stars in M67 shows that first
dredge-up modifies surface CN abundances in excess of predictions by
standard evolution calculations \citep{gilroy91,brown87}. So almost
any amount of mass transfer by a giant is likely to modify surface
abundances of blue stragglers formed in this way. Other species like
Na might also be affected \citep[see][and references
therein]{pinsonneault97}.  The straggler's companion (the original
primary) is likely to become either a helium or CO white dwarf as a
result.

This is an improvement over lithium and other light elements in that
it is a pattern change and not a destruction of the observable
species.  If the companion's contribution to the spectrum can be removed,
this could finally provide a means of determining the formation path
a blue straggler took.

\section{Conclusions}

Our spectroscopic analysis of 4 relatively cool blue stragglers does
not provide definitive evidence of composition modifications resulting
from the formation process. Based on lithium abundance patterns as a
function of effective temperature for main sequence stars in the
Hyades and M67, we find that current upper limits for the lithium
abundances of all spectroscopically observed blue stragglers are
consistent with no mixing during formation, even though the upper
limits fall below measured abundances for turnoff stars.  Although the
observable O lines are formed under NLTE conditions, the O I triplet
equivalent widths for stars in our sample and those of
\citet{mathys91} are consistent with constant abundance. The three
exceptions (S 968, S 975, and S 1434) may be caused by emission from
or absorption by other sources in a binary system.  Four of the five
blue stragglers and all of the main sequence stars have projected
rotational speeds of less than 20 km s$^{-1}$, while the fifth
straggler (S 975) has $v \sin i \approx 50$~ km s$^{-1}$.

Although we are unable to determine the existence of binary companions
to some of the blue stragglers in our sample, we are able to provide
arguments about the degree of contamination of the spectra of most of
the stragglers.  We confirm the spectroscopic detection
of a binary companion to the straggler S 1082.  From our spectra, we
measure a projected rotational speed of $90\pm20$~km s$^{-1}$ for the
secondary, and find that its radial velocity amplitude is at least 25
km s$^{-1}$. The lack of a pattern in the radial velocity data (and
the variability on the order of hours) provides additional evidence
that this system is currently undergoing mass transfer. The primary is
found to be 100 -- 200 K hotter than its companion. S 1082 may in the
end provide us with definitive proof that binary mass transfer can
produce blue stragglers.

\acknowledgments

We would like to thank the anonymous referee for a number of 
very useful comments that improved the discussion in this paper.
E.L.S. would like to thank R. Taam for support (under NSF grant
AST-9415423) and R. Taam and P. Etzel for helpful conversations during
the course of this work.

\clearpage
 
\begin{figure}
\plotone{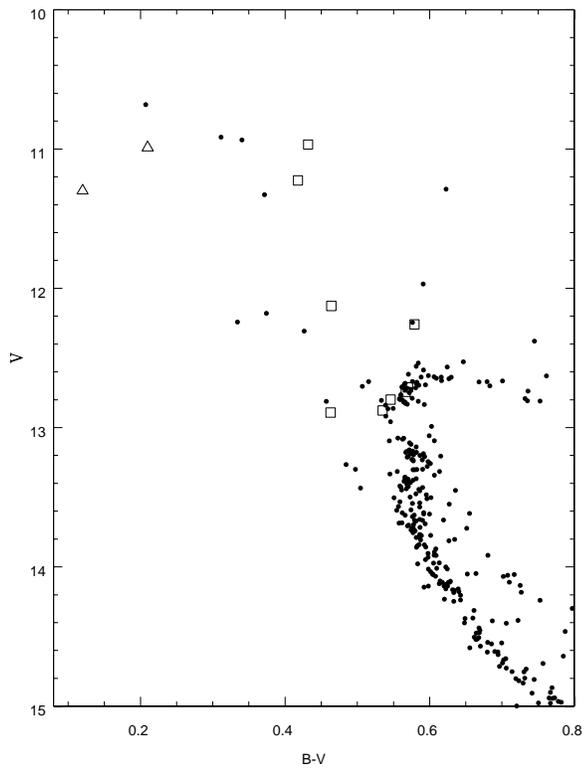}
\caption{The \citet{fan96} data is presented here with our program
stars marked as open boxes.  The open triangles are the two stars
analyzed by \citet{mathys91}.
\label{fig1}}
\end{figure}

\begin{figure}
\plotone{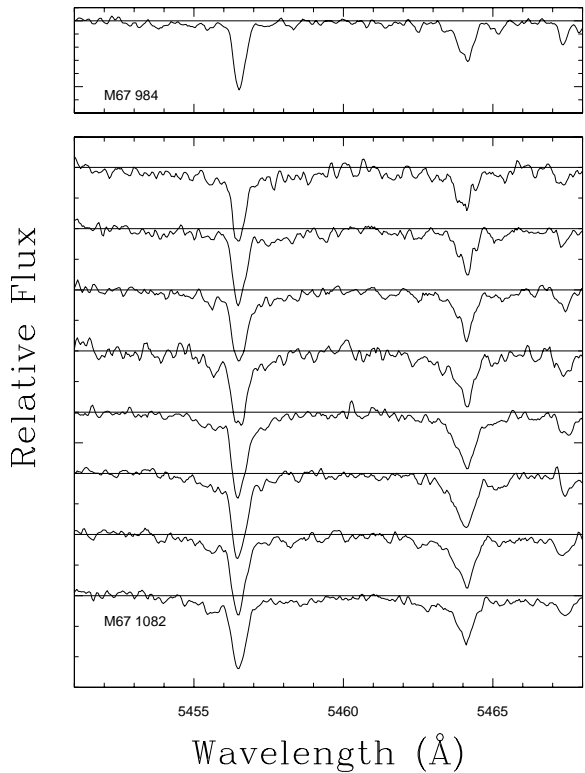}
\caption{Spectra from JD 2451619 for S 1082 and S 984 are shown
uncorrected for radial velocity variations.  
A line is drawn at the continuum for each spectrum.    The three
narrow lines are due to Fe I.  
A broad component in S 1082 can be seen in each 
spectrum but not in S 984.   The broad component of S 1082 moves from
blue to red over the 5 hours of spectra persented here (bottom to top).
\label{fig2}}
\end{figure}

\clearpage
 
\begin{figure}
\plotone{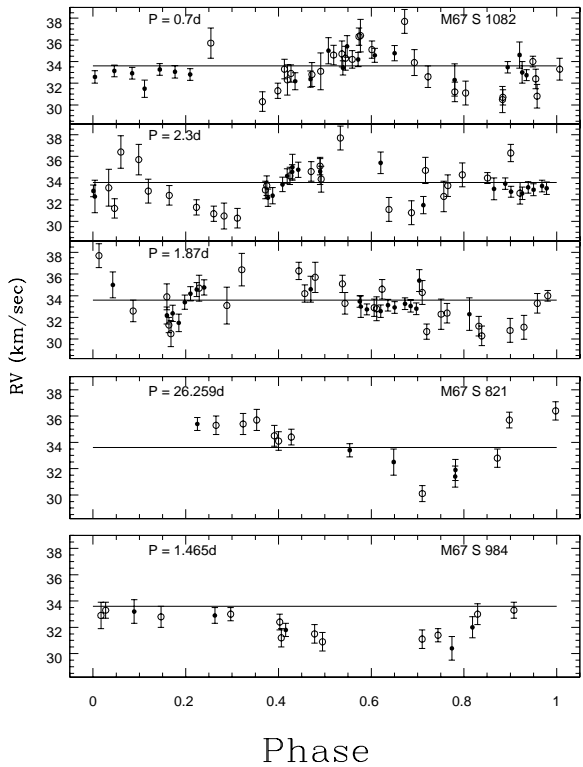}
\caption{The phased rv curves for the three detected binaries M67
S 1082, S 984, and S 821.  The open points represent data from \citet{mathieu86}
while the filled symbols represent velocities from this work.   
Phased curves of three periods are presented for S 1082.
The horizontal line represents the mean cluster velocity \citep{girard89}.
\label{fig3}}
\end{figure}

\begin{figure}
\plotone{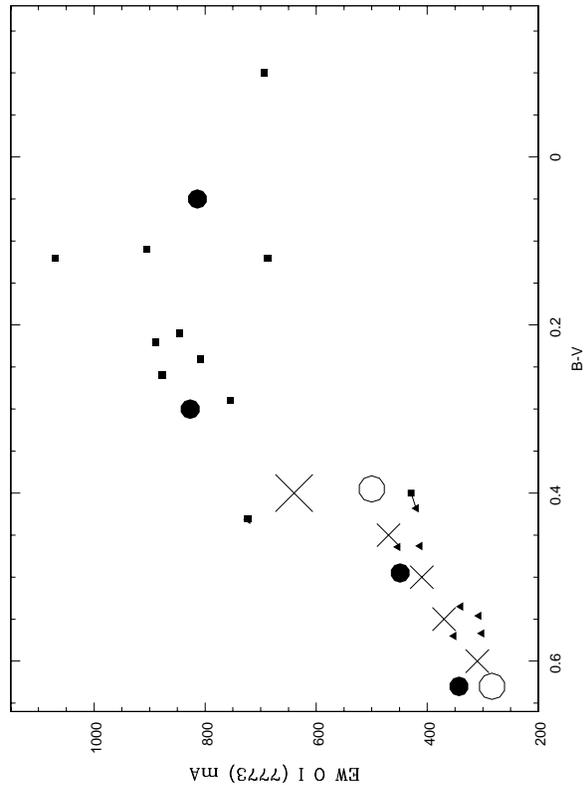}
\caption{The combined equivalent widths (EW) of the O I 7773 triplet
taken from Mathys (1991; filled squares) and this study (filled
triangles) are plotted against B-V color.  Lines connect the stars in
common between the two samples. The large crosses are taken from the
\citet{varenne99} study of oxygen in Hyades main sequence stars.  The
open and filled circles represent abundance analyses with constant
oxygen abundance ([O/Fe] = 0.0) for several temperatures using the
NLTE corrections of \citet{gratton99} and \citet{faraggiana88},
respectively.
\label{fig4}}
\end{figure}

\clearpage

\begin{deluxetable}{lrrrrr}
\tablewidth{0pt}
\scriptsize
\tablecaption{Program Stars}
\tablehead{
\colhead{star}&
\colhead{$V$}&
\colhead{$(B-V)$}&
\colhead{S/N}&
\colhead{$v \sin i$}&
\colhead{Notes}
}
\startdata
\multicolumn{6}{c}{M 67}\\
975     &   11.078  &  0.432  & 72 & 48 & BS + WD$?$ \\
1082    &   11.226  &  0.418  & 74 & 14 & BS \\
1082b   &           &         &    & 90 &  \\
997     &   12.127  &  0.464  & 69 & 16 &BS + $?$ \\
984     &   12.259  &  0.579  & 75 &$<$9&TO + TO \\
1183    &   12.711  &  0.570  & 35 &$<$9&TO \\
821     &   12.744  &  0.567  & 69 &$<$9&TO + $?$ \\
815     &   12.799  &  0.546  & 61 &$<$9&TO \\
1271    &   12.877  &  0.535  & 69 &  9 &TO \\
2204    &   12.892  &  0.463  & 44 &$<$9&BS \\
\multicolumn{6}{c}{NGC 7789}\\
1251    &   12.76   &  0.60   & 47 & 75 & NM \\
\enddata
\label{tbl-1}
\end{deluxetable}

\begin{deluxetable}{lrrrr}
\tablewidth{0pt}
\scriptsize
\tablecaption{Program Star Radial Velocities}
\tablehead{
\colhead{star}&
\colhead{HJD}&
\colhead{Telescope}&
\colhead{rv (km s$^{-1}$)}&
\colhead{$\sigma$(rv)}
}
\startdata
M67 S 975    &   2450828.72  &  107  & 28.5 & 1.5 \\
             &   2450853.97  &  107  & 33.0 & 4.3 \\
             &   2451243.79  &  HET  & 37.8 & 2.3 \\
             &   2451567.90  &  HET  & 37.5 & 1.1 \\
M67 MMJ5554  &   2451567.91  &  HET  & 30.0 & 1.1 \\
M67 S 1082a  &   2450828.88  &  107  & 31.5 & 0.8 \\
             &   2450855.62  &  107  & 34.6 & 1.2 \\
             &   2451258.75  &  HET  & 32.3 & 1.5 \\
             &   2451559.94  &  HET  & 35.4 & 1.0 \\
             &   2451554.96  &  HET  & 35.0 & 1.2 \\
             &   2451555.96  &  HET  & 33.0 & 1.0 \\
             &   2451619.60  &   82  & 33.5 & 0.6 \\
             &   2451619.63  &   82  & 32.7 & 0.6 \\
             &   2451619.69  &   82  & 32.6 & 0.7 \\
             &   2451619.72  &   82  & 33.1 & 0.6 \\
             &   2451619.74  &   82  & 32.9 & 0.6 \\
             &   2451619.79  &   82  & 33.3 & 0.6 \\
             &   2451619.81  &   82  & 33.1 & 0.7 \\
             &   2451619.83  &   82  & 32.8 & 0.6 \\
             &   2451620.70  &   82  & 32.2 & 0.9 \\
             &   2451620.72  &   82  & 32.4 & 0.9 \\
             &   2451620.77  &   82  & 33.4 & 0.8 \\
             &   2451620.79  &   82  & 34.2 & 0.8 \\
             &   2451620.82  &   82  & 34.6 & 0.7 \\
             &   2451620.85  &   82  & 34.8 & 0.8 \\
M67 S 1082b  &   2450828.88  &  107  & 54   & 6   \\
             &   2450855.62  &  107  & 31   & 10  \\
             &   2451619.60  &   82  & 30   & 6   \\
             &   2451619.63  &   82  & 32   & 9   \\
             &   2451619.69  &   82  & 43   & 7   \\
             &   2451619.72  &   82  & 34   & 9   \\
             &   2451619.74  &   82  & 30   & 7   \\
             &   2451619.79  &   82  & 51   & 12  \\
             &   2451619.81  &   82  & 53   & 7   \\
             &   2451619.83  &   82  & 47   & 11  \\
             &   2451620.70  &   82  & 37   & 12  \\
M67 S 997    &   2450828.93  &  107  & 30.1 & 0.5 \\
             &   2451239.23  &  107  & 29.0 & 0.7 \\
             &   2451568.74  &  HET  & 30.0 & 1.1 \\
M67 S 984    &   2450853.69  &  107  & 31.8 & 0.5 \\
             &   2451554.95  &  HET  & 33.2 & 0.9 \\
             &   2451555.95  &  HET  & 30.4 & 0.9 \\
             &   2451558.94  &  HET  & 32.0 & 0.8 \\
             &   2451619.66  &   82  & 32.9 & 0.6 \\
M67 S 1183   &   2450855.83  &  107  & 33.3 & 0.5 \\
M67 S 821    &   2450853.90  &  107  & 33.4 & 0.5 \\
             &   2451239.17  &  107  & 35.4 & 0.5 \\
             &   2451568.89  &  HET  & 31.4 & 0.8 \\
             &   2451568.91  &  HET  & 31.9 & 0.8 \\
             &   2451591.67  &  HET  & 32.5 & 1.0 \\
M67 S 815    &   2450855.74  &  107  & 33.2 & 0.5 \\
M67 S 1271   &   2450853.78  &  107  & 34.0 & 0.7 \\
             &   2450854.82  &  107  & 33.5 & 0.9 \\
             &   2450855.65  &  107  & 34.3 & 0.6 \\
M67 S 2204   &   2450828.78  &  107  & 34.2 & 0.6 \\
             &   2450855.84  &  107  & 33.6 & 0.8 \\
             &   2451554.94  &  HET  & 33.8 & 1.1 \\
             &   2451555.93  &  HET  & 32.8 & 0.8 \\
             &   2451558.93  &  HET  & 32.1 & 0.9 \\
NGC7789 1251 &   2450853.60  &  107  & -16.4 & 4.0 \\
             &   2450855.57  &  107  & -16.7 & 5.0 \\
\enddata
\label{tbl-2}
\end{deluxetable}

\begin{deluxetable}{lrrrrrrrrrr}
\tablewidth{0pt}
\scriptsize
\tablecaption{EW of Program Stars}
\tablehead{
\colhead{$\lambda$}&
\colhead{gf}&
\colhead{975}&
\colhead{1082}&
\colhead{997}&
\colhead{984}&
\colhead{1183}&
\colhead{821}&
\colhead{815}&
\colhead{1271}&
\colhead{2204}
}
\startdata
\multicolumn{11}{c}{Li I}\\
  6707.8   &  0.17 & $<$15   & $<$2.0  & $<$2.0  &  25     &  $<$10  & $<$10   & $<$7    & $<$6    & $<$5    \\
\multicolumn{11}{c}{C I}\\                                                                                  \\
  5380.34  & -1.62 & 57      & 72      & 48      &  28     & 45      &  38     &  39     &  45     &  51     \\
  7111.48  & -1.09 & \nodata & 23      & 17      & \nodata & \nodata &  25     &  19     &  28     &  27     \\
  7113.18  & -0.77 & 43      & 42      & 47      &  40     & 39      &  24     &  38     &  39     &  36     \\
  7115.19  & -0.93 & 49      & 42      & 37      &  35     & 30      &  41     &  27     &  42     &  48     \\
  7119.67  & -1.15 & \nodata & 25      & 38      &  28     & 25      &  11     &  20     &  21     &  29     \\
\multicolumn{11}{c}{N I}\\                                                                                  
  7468.31  & -0.19 & $<$17   & $<$8    &  8      &  12     &  9      &  10     &  $<$12  & $<$8    & $<$10   \\
\multicolumn{11}{c}{O I}\\                                                                                  
  7771.94  &  0.37 & 296     & 145     & 169     & 106     & 138     & 115     & 115     & 126     & 158     \\
  7774.17  &  0.22 & 238     & 145     & 151     & \nodata & 131     &  95     & 106     & 130     & 142     \\
  7775.39  &  0.00 & 192     & 130     & 133     &  84     &  83     &  92     &  86     &  84     & 113     \\
\multicolumn{11}{c}{Na I}\\                                                                                 
  6154.23  & -1.56 & \nodata & \nodata &  14     &  24     &  19     &  26     &  29     &  24     &  15     \\
  6160.75  & -1.26 & 10      & 24      &  38     &  47     &  43     &  39     &  38     &  29     &  35     \\
\multicolumn{11}{c}{Mg I}\\                                                                                 
  5528.40  & -0.56 & 195     & 87      & 215     & 199     & 160     & 190     & 196     & 184     & 178     \\
  5711.10  & -1.73 &  75     & 36      &  68     &  92     &  94     &  95     &  92     &  84     &  63     \\
\multicolumn{11}{c}{Ca I}\\                                                                                 
  5588.75  &  0.21 & \nodata & \nodata & \nodata & 149     & 130     & 150     & 180     & 133     & 124     \\
  5857.45  &  0.23 & 145     & 70      & 140     & 117     & 126     & 120     & 126     & 124     & 120     \\
  6161.30  & -1.27 & 67      & 23      &  38     &  58     &  47     &  43     &  59     &  52     &  41     \\
  6162.17  & -0.09 & 210     & 75      & 167     & 164     & 163     & 168     & 172     & 169     & 142     \\
  6166.44  & -1.14 & 50      & 23      &  36     &  67     &  89     &  56     &  58     &  56     &  53     \\
  6169.04  & -0.80 & 86      & 42      &  67     &  83     &  88     &  84     &  72     &  78     &  60     \\
  6169.56  & -0.48 & 118     & \nodata &  83     & 108     & 107     &  98     &  92     &  95     &  83     \\
  6439.08  &  0.19 & 206     & 87      & 142     & 153     & 162     & 151     & 144     & 148     & 137     \\
  6455.60  & -1.29 & 64      & \nodata &  32     &  54     &  50     &  51     &  45     &  47     &  25     \\
  6499.65  & -0.82 & 68      & 27      &  75     &  75     &  80     &  75     &  68     &  73     &  56     \\
\multicolumn{11}{c}{Fe I}\\                                                                                 
  5060.08  & -5.06 & \nodata & 9       &  20     &  55     &  28     &  57     &  39     &  29     &  24     \\
  5067.16  & -0.93 & \nodata & 32      &  49     &  63     &  66     &  59     &  43     &  51     &  39     \\
  5083.34  & -3.31 & \nodata & 37      &  88     & 103     &  99     & 100     & 103     &  90     &  82     \\
  5088.16  & -1.59 & \nodata & \nodata &  19     &  35     & \nodata & \nodata & \nodata & \nodata & \nodata \\
  5090.78  & -0.60 & 70      & 37      &  80     &  87     &  75     &  83     &  65     &  65     &  80     \\
  5109.66  & -0.80 & 60      & 32      &  53     &  63     &  84     &  54     &  81     &  74     &  53     \\
  5126.20  & -0.85 & \nodata & 25      &  50     &  62     &  52     &  73     &  55     &  47     & \nodata \\
  5127.37  & -3.52 & 60      & 29      &  70     &  88     &  79     &  89     &  75     &  75     &  71     \\
  5141.75  & -2.25 & 70      & 22      &  66     & \nodata &  76     &  76     &  83     &  84     &  73     \\
  5225.53  & -4.79 & \nodata &  7      &  17     &  61     &  66     &  56     &  48     &  37     & \nodata \\
  5809.22  & -1.76 & 25      & 13      &  24     &  41     &  33     &  42     &  34     &  28     &  25     \\
  5852.22  & -1.22 & \nodata & \nodata &  22     &  35     &  35     &  30     &  30     &  25     & \nodata \\
  5855.08  & -1.56 & \nodata &  6      &   9     &  19     & \nodata &  16     &  16     &   9     &  13     \\
  5856.09  & -1.60 & \nodata & \nodata &  23     & \nodata & \nodata &  33     & \nodata & \nodata & \nodata \\
  5859.59  & -0.61 & 60      & 25      &  72     &  63     &  63     &  71     &  61     &  52     &  62     \\
  5862.36  & -0.42 & \nodata & \nodata &  69     &  85     & \nodata &  80     &  75     &  76     & \nodata \\
  6151.62  & -3.35 & 15      & \nodata &  20     &  45     &  30     &  28     &  28     &  28     &  16     \\
  6157.73  & -1.25 & 50      & 17      &  30     &  60     &  46     &  50     &  48     &  41     &  52     \\
  6165.37  & -1.50 & \nodata & \nodata &  15     &  42     &  38     &  38     &  29     &  30     &  16     \\
  6173.34  & -2.90 & 40      & 17      &  30     &  71     &  62     &  63     &  56     &  42     &  35     \\
  6187.99  & -1.70 & \nodata & \nodata & \nodata &  46     &  24     &  53     &  32     &  30     &  25     \\
  6213.43  & -2.66 & 55      & 19      &  50     & \nodata &  94     &  83     &  65     &  71     &  64     \\
  6219.28  & -2.50 & 70      & 28      &  50     &  81     &  91     &  81     &  72     &  76     &  64     \\
  6226.74  & -2.22 & \nodata & \nodata & \nodata &  20     & \nodata &  24     &  16     &  17     &  14     \\
  6229.23  & -3.00 & \nodata & \nodata &  16     &  26     &  32     &  30     &  31     &  22     & \nodata \\
  6240.65  & -3.23 & \nodata & \nodata & \nodata &  43     & \nodata &  30     &  33     &  24     & \nodata \\
  6280.62  & -4.37 & \nodata & \nodata & \nodata & \nodata &  54     &  44     &  43     &  30     &  24     \\
  6290.97  & -0.76 & \nodata & 17      & \nodata &  58     & \nodata &  67     &  45     &  52     &  32     \\
  6297.79  & -2.80 & \nodata & \nodata & \nodata & \nodata &  80     & \nodata &  87     &  70     &  64     \\
  6301.50  & -0.80 & \nodata & \nodata & \nodata & \nodata & \nodata & \nodata &  79     &  84     &  94     \\
  6355.03  & -2.29 & 40      & 21      &  40     &  66     &  67     &  55     &  59     &  49     &  50     \\
  6380.74  & -1.40 & 24      & 10      &  20     &  46     &  55     &  41     &  42     &  33     &  30     \\
  6498.94  & -4.69 & \nodata & \nodata &  17     &  24     & \nodata &  25     &  23     &  18     & \nodata \\
  7802.51  & -1.37 & \nodata & \nodata & \nodata &  14     &  18     &  11     & \nodata &  11     & \nodata \\
  8757.19  & -2.02 & \nodata & \nodata & \nodata & \nodata & \nodata &  86     &  84     &  64     &  64     \\
\multicolumn{11}{c}{Fe II}\\                                                                                
  4923.92  & -1.44 & 230     & 148     & 226     & 163     & 179     & 173     & 168     & 165     & 181     \\
  5100.66  & -4.13 & \nodata &  19     &  44     &  27     & \nodata &  44     & \nodata &  37     & \nodata \\
  5264.79  & -3.23 & 70      &  33     &  57     &  73     & \nodata &  57     &  56     &  47     &  64     \\
  5414.08  & -3.48 & \nodata &  20     &  30     &  50     &  45     &  43     &  34     &  35     &  33     \\
  6149.23  & -2.76 & 70      &  40     &  65     &  60     &  44     &  55     &  61     &  49     &  36     \\
  6369.46  & -4.25 & 23      &  10     &  19     &  35     &  29     &  39     &  33     &  28     &  23     \\
  6416.92  & -2.79 & \nodata & \nodata & \nodata &  58     & \nodata & \nodata &  50     &  44     &  67     \\
  6456.39  & -2.08 & 120     &  76     & 110     &  97     &  88     &  94     &  88     &  75     &  88     \\
  6516.08  & -3.45 & 90      &  58     &  80     &  87     &  77     &  87     &  75     &  85     &  81     \\
  7449.34  & -3.06 & 40      &  23     &  20     &  34     &  38     &  30     &  30     &  38     &  26     \\
\multicolumn{11}{c}{Ni I}\\                                                                                 
  6175.37  & -0.53 & \nodata &  18     &  26     &  33     &  39     &  33     &  59     &  46     &  41     \\
  6176.82  & -0.53 & \nodata &  24     &  42     &  56     &  43     &  54     &  48     &  43     &  63     \\
  6378.26  & -0.89 & \nodata &   8     &  23     &  26     &  16     &  23     &  17     &  20     &  16     \\
  7122.19  & -0.16 & 75      & \nodata &  83     & 100     & 110     &  91     &  90     &  74     &  82     \\
\multicolumn{11}{c}{Ba I}\\                                                                                 
  5853.68  & -1.01 & 94      &  34     &  74     &  83     & 117     &  82     &  71     &  85     &  78     \\
  6141.73  & -0.08 & 200     &  71     & 145     & 137     & 140     & 128     & 127     & 125     & 130     \\
  6496.91  & -0.38 & 121     &  69     & 150     & 134     & 152     & 126     & 125     & 126     & 122     \\
\enddata
\label{tbl-3}
\tablecomments{Li EW limits quoted here were taken from 
\citet{pritchet91} for 1082 and 997 because their limits were smaller than ours.}
\end{deluxetable}

\begin{deluxetable}{lrrrrrllrrrlc}
\rotate
\tablewidth{0pt}
\tablecaption{Program Star Abundances}
\tablehead{
\colhead{star}&
\colhead{$T_{eff}$}&
\colhead{g}&
\colhead{[Fe/H]}&
\colhead{Li}&
\colhead{[C/Fe]}&
\colhead{[N/Fe]}&
\colhead{[Na/Fe]}&
\colhead{[Mg/Fe]}&
\colhead{[Ca/Fe]}&
\colhead{[Ni/Fe]}&
\colhead{[Ba/Fe]}&
\colhead{Note} 
}
\startdata
\multicolumn{13}{c}{\underline{Blue Stragglers}} \\
\multicolumn{13}{l}{Our Data:} \\
\multicolumn{13}{l}{Probably Minimal or No Contamination by Companions:}\\
984     & 6170 & 3.9 & $0.08\pm0.03$ &  \phn 2.20:  & $-0.14\pm0.08$ & \phn$0.10\pm0.2$    & $-0.13\pm0.06$ & $-0.09\pm0.06$ & $-0.05\pm0.03$ & $-0.01\pm0.09$ &  \phn$0.25\pm0.06$ & \\
2204    & 6650 & 4.6 & $-0.05\pm0.06$ & $<$2.00 &  $0.05\pm0.06$ & $<$0.0    & $-0.05\pm0.06$ & $-0.26\pm0.06$ & $-0.17\pm0.05$ &  $0.17\pm0.14$ &  \phn$0.12\pm0.05$ & \\
\multicolumn{13}{l}{Unknown Contamination by Companion:}\\
975     & 6820 & 4.4 & $0.02\pm0.06$ & $<$2.50 & $-0.09\pm0.09$ & $<$0.05 & $-0.6$:        & $-0.13\pm0.07$ &  $0.19\pm0.05$ & $-0.25\pm0.20$ &  \phn$0.23\pm0.21$ & cb\\
997     & 6675 & 4.4 & $-0.06\pm0.03$ & $<$1.60 & $-0.03\pm0.08$ & $-0.15\pm0.2$    & $-0.02\pm0.11$ & $-0.04\pm0.13$ & $-0.04\pm0.03$ &  $0.10\pm0.09$ &  \phn$0.22\pm0.11$ & eb \\
\multicolumn{13}{l}{Sure or Probable Contamination by Companion:}\\
1082    & 7050 & 4.5 & $-0.25\pm0.05$ & $<$1.80 & $-0.01\pm0.05$ & $<$0.0    & \phn$0.23\pm0.20$  & $-0.41\pm0.18$ & $-0.12\pm0.07$ &  $0.23\pm0.10$ & $-0.02\pm0.05$ & acc? \\
\multicolumn{13}{l}{Mathys (1991) Data:} \\
968     & 8560 & 4.1 & $0.03\pm0.16$ &          & $-0.75\pm0.23$ & $-0.16\pm0.21$   & \phn$0.10$  & $-0.38\pm0.17$ & $-0.70\pm0.18$ & $0.23\pm0.17$ & \phn$1.06$ & Am\\
1263    & 8290 & 4.1 & $0.09\pm0.19$ &          & $-0.64\pm0.21$ & $-0.07$   & $-0.11$        & $-0.30\pm0.19$ & $-0.22\pm0.25$ & $0.33\pm0.22$ & \phn$0.76$ & \\
\multicolumn{13}{c}{\underline{Turnoff Stars}} \\
815     & 6275 & 4.2 & $-0.05\pm0.04$ & $<$1.90 & $-0.05\pm0.04$ & $<$0.15 &  \phn$0.01\pm0.07$ & $-0.05\pm0.07$ & $-0.03\pm0.04$ &  $0.10\pm0.12$ &  \phn$0.11\pm0.07$ & \\
1183    & 6250 & 4.2 & $-0.04\pm0.08$ & $<$2.00 &  $0.05\pm0.05$ & \phn$0.15\pm0.2$  & $-0.10\pm0.08$ & $-0.23\pm0.22$ & $-0.04\pm0.06$ &  $0.00\pm0.05$ &  \phn$0.43\pm0.16$ & \\
1271    & 6360 & 4.3 & $-0.07\pm0.03$ & $<$1.90 &  $0.10\pm0.04$ & $<$0.05 & $-0.07\pm0.10$ & $-0.08\pm0.07$ &  $0.02\pm0.03$ &  $0.06\pm0.10$ &  \phn$0.32\pm0.06$ & \\
 & & & & & & & & & & & & \\
821     & 6190 & 4.0 & $-0.04\pm0.04$ & $<$2.00 & $-0.12\pm0.10$ & \phn$0.20\pm0.2$  & $-0.05\pm0.06$ & $-0.08\pm0.08$ & $-0.01\pm0.03$ & $-0.02\pm0.09$ & $-0.02\pm0.06$ & eb \\
\enddata
\tablecomments{eb --- eccentric binary; cb --- circularized
binary; acc? --- accretion occurring in binary?; Am --- known Am star}
\label{tbl-4}
\end{deluxetable}

\begin{deluxetable}{lrrrrrrr}
\tablewidth{0pt}
\scriptsize
\tablecaption{Program Star Oxygen Abundances}
\tablehead{
\colhead{star}&
\colhead{T$_{eff}$}&
\colhead{g}&
\colhead{v$_t$}&
\colhead{[Fe/H]}&
\colhead{[O/Fe]}&
\colhead{correction}&
\colhead{[O/Fe]$_{NLTE}$}
}
\startdata
\multicolumn{8}{c}{Blue Stragglers} \\
975     & 6820 & 4.4 & 3.1 &  0.02(0.06) &  0.70+/-0.08 & -0.26 &  0.44 \\
997     & 6675 & 4.4 & 2.6 & -0.06(0.03) &  0.24+/-0.05 & -0.24 &  0.00 \\
1082    & 7050 & 4.5 & 1.6 & -0.25(0.05) &  0.30+/-0.06 & -0.30 &  0.16 \\
2204    & 6650 & 4.6 & 2.5 & -0.05(0.06) &  0.15+/-0.05 & -0.24 & -0.09 \\
\multicolumn{8}{c}{Turnoff Stars} \\
821     & 6190 & 4.0 & 2.0 & -0.04(0.04) &  0.01+/-0.05 & -0.09 & -0.08 \\
1183    & 6250 & 4.2 & 2.0 & -0.04(0.08) &  0.16+/-0.10 & -0.10 &  0.06 \\
815     & 6275 & 4.2 & 1.8 & -0.05(0.04) &  0.02+/-0.05 & -0.10 & -0.08 \\
1271    & 6360 & 4.3 & 1.8 & -0.07(0.03) &  0.15+/-0.09 & -0.11 &  0.04 \\
984     & 6170 & 3.9 & 1.5 &  0.08(0.03) & -0.13+/-0.06 & -0.09 & -0.22 \\
\enddata
\label{tbl-5}
\end{deluxetable}
\end{document}